\documentclass[letterpaper, 10 pt, conference]{ieeeconf}

\IEEEoverridecommandlockouts                              %
\usepackage{amsmath,amssymb,amsfonts}
\usepackage{mathtools}

\usepackage{xcolor}
\usepackage{tikz,pgfplots}
\usepackage{bbding} %
\usepackage{ifthen} %

\colorlet{darkGreen}{green!80!black}
\colorlet{lightblue}{blue!10!white}
\colorlet{darkOrange}{orange!90!black}
\colorlet{checkmarkGreen}{green!80!black}

\definecolor{istblue}{rgb/cmyk}{0.0,0.25490,0.56862745/1,0.55,0.0,0.43}%
\definecolor{istdarkgreen}{rgb/cmyk}{0.0627,0.5882,0.2824/0.89,0.0,0.52,0.41}%

\usetikzlibrary{arrows,arrows.meta,shapes,positioning,calc,decorations.pathreplacing,calligraphy}

\tikzstyle{smallCircle} = [draw, circle, fill=black,inner sep=0pt,outer sep=0.7pt,minimum size=2pt]
\tikzstyle{block} = [draw, rectangle,minimum height=3em, minimum width=3em]
\tikzstyle{sum} = [draw, circle]
\tikzstyle{box}=[rectangle, fill=gray!20, draw, minimum width=1.2cm, minimum height=0.5cm, align=center]
\tikzset{vertex/.style = {shape=circle,draw,minimum size=1.5em,inner sep=0}}
\tikzset{vertexFill/.style = {shape=circle,fill=gray!40!white,draw,minimum size=1.5em,inner sep=0}}
\tikzset{vertexFillMarked/.style = {shape=circle,fill=orange!50!white,draw,minimum size=1.5em,inner sep=0}}
\tikzset{edge/.style = {->,> = latex'}}
\tikzset{edgeMarked/.style = {->,> = latex',thick,darkOrange}}
\tikzset{lightning bolt to/.style={to path={
			let \p1=(\tikztostart), \p2=(\tikztotarget), \n1={veclen(\y2-\y1,\x2-\x1)} in
			(\p1) -- ($($(\p1)!0.6!(\p2)$)!\n1*.1!-90:(\p2)$) -- ($(\p1)!0.55!(\p2)$) --
			(\p2) -- ($($(\p1)!0.4!(\p2)$)!\n1*.1!90:(\p2)$) -- ($(\p1)!0.45!(\p2)$) -- 
			cycle (\p2)%
}}}

\newcommand{\norm}[1]{\left\lVert#1\right\rVert}
\newcommand{\ie}{i.e., }

\newcommand{\cf}{cf.\ }

\newtheorem{remark}{Remark}
\newtheorem{theorem}{Theorem}
\newtheorem{corollary}{Corollary}
\newtheorem{definition}{Definition}

\usepackage{microtype}
\let\labelindent\relax %
\usepackage{enumitem}
\usepackage[group-separator={,}]{siunitx}

\usepackage{textcomp,hyperref}
\newcommand\copyrighttext{%
  \footnotesize \textcopyright 2025 IEEE. Personal use of this material is permitted.
  Permission from IEEE must be obtained for all other uses, in any current or future 
  media, including reprinting/republishing this material for advertising or promotional 
  purposes, creating new collective works, for resale or redistribution to servers or 
  lists, or reuse of any copyrighted component of this work in other works. 
  }
\newcommand\copyrightnotice{%
\begin{tikzpicture}[remember picture,overlay]
\node[anchor=south,yshift=10pt] at (current page.south) {\fbox{\parbox{\dimexpr\textwidth-\fboxsep-\fboxrule\relax}{\copyrighttext}}};
\end{tikzpicture}%
}

\usepackage{nccmath}
\usepackage{xparse,etoolbox}
\DeclarePairedDelimiterX{\innerp}[1]{\langle}{\rangle}{\innpargs{#1}}
\NewDocumentCommand{\innpargs}{>{\SplitArgument{1}{,}}m}
{\innpargsaux#1}
\NewDocumentCommand{\innpargsaux}{mm}
{\ifblank{#1}%
	{\ifblank{#2}{~{,}~}{{\,\cdot\,}{,}{\mkern2mu#2}}}%
	{{#1\,}{,}\ifblank{#2}{\,\cdot\,}{\mkern2mu#2}}%
}%

\newcommand{\nv}{{n_\mathcal{V}}}
\newcommand{\K}{{\mathcal{K}}}
\newcommand{\interior}[1]{\mathrm{int}(#1)}

\title{\LARGE \bf
Performance analysis for cone-preserving switched systems with constrained switching
}

\author{Marc Seidel, Richard Pates, Frank Allgöwer%
\thanks{F.~Allg\"ower and M.~Seidel thank the German Research Foundation (DFG) for support of this work within grant AL 316/13-2 and within the German Excellence Strategy under grant EXC-2075 - 285825138; 390740016.}%
\thanks{R.~Pates thanks the ELLIIT Strategic Research Area at Lund University for support. He additionally gratefully acknowledges the initiative COMPEL, with funding from the Swedish government, as well as the ERC through grant agreement No. 834142.}
\thanks{M.~Seidel and F.~Allgöwer are with the University of Stuttgart, Institute for Systems Theory and Automatic Control, Stuttgart, Germany, {\tt \small\{seidel,allgower\}@ist.uni-stuttgart.de}. R.~Pates is with the Department of Automatic Control LTH, Lund University, Box 118, SE-
221 00 Lund, Sweden {\tt \small richard.pates@control.lth.se}.}%
}

\begin{document}

\maketitle
\copyrightnotice
\thispagestyle{empty}
\pagestyle{empty}

\setlength{\abovedisplayshortskip}{1ex plus1ex minus1ex}
\setlength{\abovedisplayskip}{1ex plus1ex minus1ex}
\setlength{\belowdisplayshortskip}{1ex plus1ex minus1ex}
\setlength{\belowdisplayskip}{1ex plus1ex minus1ex}

\begin{abstract}
    This paper studies cone-preserving linear discrete-time switched systems whose switching is governed by an automaton.
    For this general system class, we present performance analysis conditions for a broadly usable performance measure. %
    In doing so, we generalize several known results for performance and stability analysis for switched and positive switched systems, providing a unifying perspective.
    We also arrive at novel $\ell_1$-performance analysis conditions for positive switched systems with constrained switching, for which we present an application-motivated numerical example.
    Further, the cone-preserving perspective provides insights into appropriate Lyapunov function selection.
\end{abstract}

\section{Introduction} \label{sec:introduction}
Switched systems, characterized by dynamics that switch between multiple subsystems, are prevalent in many engineering applications. %
Examples include networked control systems, automotive systems, and real-time control applications, where the system's mode may change due to component failures, varying operational demands, or scheduling protocols.
Analyzing the stability and performance of such systems is crucial, especially when the switching is governed by specific rules or constraints \cite{Lin2009}.

In this paper, we study stability and performance of switched systems for which the systems' state remains within a prescribed cone in the state space. This generalizes the well-known class of positive systems, where the state remains in the nonnegative orthant. Positive systems appear in various applications, including population dynamics, epidemic modeling, and economics, where system variables must remain non-negative \cite{Rantzer2018,Chen2013, Fornasini2012, Liu2009}. More broadly, cone-preserving systems arise in contexts where state interactions are constrained by an underlying conic structure. To date their systematic study has been relatively limited, though a range of definitions and analysis results can be found in \cite{Shen2017}.

A key challenge in analyzing such systems is the selection of appropriate Lyapunov function candidates. In general switched system analysis, Lyapunov functions based on quadratic functions are widely used due to their strong theoretical properties and computational tractability. In contrast, for positive systems, linear Lyapunov functions are commonly employed \cite{Liu2009}, offering simpler and more scalable stability and performance criteria. For cone-preserving systems, the choice of Lyapunov function is less obvious, as the cone structure imposes additional constraints on the function selection.
Some recent works have explored path-complete positivity as a possible tool for analyzing these systems \cite{Forni2017}. %
The many specialized analysis tools for each specific problem highlight the need for a structured approach for stability and performance characterization.

In addition to the cone-preserving property, the presence of switching constraints introduces further analytical challenges. In many real-world systems, switching is not arbitrary, but is instead governed by external rules, such as real-time scheduling policies or communication constraints. A prominent example are weakly-hard control systems, where switching sequences must satisfy predefined constraints due to computational or communication limitations \cite{Vreman2022b, Seidel2024b}. Stability analysis for positive switched systems, both under arbitrary and constrained switching, has been widely explored \cite{Lin2009, An2024, Benzaouia2012, Fornasini2012, Liu2009}.
For cone-preserving switched systems, some stability results exist \cite{Shen2010, Bundfuss2009}.
However, performance analysis, particularly for general cone-preserving switched systems, remains less developed, with existing results often tailored to specific performance measures.

In this work, we establish performance analysis conditions for cone-preserving linear discrete-time switched systems with switching constrained by an automaton. Our results offer a unifying perspective that includes a widely usable performance measure, generalizing existing results on positive switched systems and more general switched system performance analysis. Specifically, we show that our framework recovers well-established results from the literature, including
$\ell_2$-performance for switched systems with constrained switching \cite{Seidel2024b}, $\ell_1$-performance for positive and cone-preserving systems \cite{Chen2013, Zhu2024, Shen2017}, and stability results for positive switched systems \cite{Liu2009, Fornasini2012, Benzaouia2012}.
Furthermore, as a consequence of our analysis, we derive $\ell_1$-performance conditions for positive switched systems with constrained switching, and provide an application-motivated numerical example illustrating the applicability of our results.
Finally, using the cone-preserving perspective, we propose a structured Lyapunov function candidate for cone-preserving switched systems, providing insights into the appropriate function selection.%

The remainder of the paper is structured as follows.
Section~\ref{sec:setup} introduces the considered setup and system class. Section~\ref{sec:performance} contains the main result, followed by examples in Section~\ref{sec:examples}, showing the generality of the main result.
Section~\ref{sec:conclusion} concludes the paper.

\emph{Notation.}
The set of real-valued $n$-dimensional vectors is denoted $\mathbb{R}^n$ with the subscript $\geq{}\!\!0$, $>{}\!\!0$, or $<{}\!\!0$ specifying nonnegative, strictly positive, or strictly negative entries, respectively.
The same notation applies to $n \times m$ matrices $\mathbb{R}^{n \times m}$.
For a quadratic matrix $M$, $\mathrm{tr}(M)$ is the trace of $M$ and $M^\top$ is its transpose.
For $M$ symmetric, $M \succ 0$ and $M \prec 0$ denotes positive or negative definite, respectively.
The identity matrix is $I$ and $\mathsf{1}$ is the vector of ones of suitable dimensions.
The $p$-norm of a sequence or vector $x$ is $\norm{x}_p$.

\section{Setup} \label{sec:setup}
In this section, we introduce our considered system class, the underlying switching rule, and performance measure, that are defined on cones.
For that, we revisit some basic definitions and facts on cones first.

\subsection{Cones and operators} \label{sec:setup-cones}
In this paper, we consider finite dimensional proper cones $\K$ defined over a real vector space, \ie $\K$ is convex, closed, pointed, and has a nonempty interior \cite{Boyd2004}, denoted $\interior{\K}$.
Such cones are equipped with an inner product, that can be used to define the corresponding dual cone $\K^*$ \cite{Boyd2004} as
\begin{equation}
\begin{aligned} \label{eq:dual-cone}
	\begin{split}
		\K^* &= \{\, y \mid \innerp{y,x} \geq 0 \text{ for all } x \in \K \,\} \\
		\interior{\K^*} &= \{\, y \mid \innerp{y,x} > 0 \text{ for all } x \in \K  \,\}.
	\end{split}
\end{aligned}
\end{equation}
A cone is called \emph{self-dual} if $\K = \K^*$.
Further, an \emph{operator} $\mathbf{A}$ from a cone $\K_x$ to another cone $\K_y$ is a mapping $\mathbf{A} \colon \K_x \to \K_y$.
Throughout the paper, we use boldfont letters to distinguish operators from matrices or vectors, and calligraphic letters for sets.
Moreover, for each linear operator $\mathbf{A}$, there exists an \emph{(Hermitian) adjoint} operator $\mathbf{A}^* \colon \K_y^* \to \K_x^*$ defined by $\innerp{\mathbf{A}^* y,x} = \innerp{y,\mathbf{A}x}$, where $x \in \K_x$, $y \in \K_y^*$.
For a vector of operators
\begin{equation}
\begin{aligned}
	\begin{split} \label{eq:adjoint-2elements}
		\innerp*{
			\begin{pmatrix} \mathbf{A} & \mathbf{B} \end{pmatrix}^* y,
			\begin{pmatrix} x \\ w \end{pmatrix}
		} &= \innerp*{
			\begin{pmatrix} \mathbf{A}^* y \\ \mathbf{B}^* y \end{pmatrix},
			\begin{pmatrix} x \\ w \end{pmatrix}
		} \\
		= \innerp*{
			y,
			\begin{pmatrix} \mathbf{A} & \mathbf{B} \end{pmatrix} \begin{pmatrix} x \\ w \end{pmatrix}
		} &= \innerp*{
			y,
			\mathbf{A} x + \mathbf{B} w
		}
	\end{split}
\end{aligned}
\end{equation}
with $w \in \K_w$, $\mathbf{B} \colon \K_w \to \K_y$, and $\mathbf{B}^* \colon \K_y^* \to \K_w^*$. %
The in the next subsection formally defined switched system will admit dynamics on a cone $\K$, and its dual is used to define a performance measure and Lyapunov function thereafter.

\subsection{Cone-preserving switched systems} \label{sec:setup-cone-systems}
Consider two proper cones and elements thereof, $\begin{pmatrix} x(t) \\ w(t) \end{pmatrix} \in \K$ and $z(t) \in \K_z$.
Further, we denote $\K_x$ and $\K_w$ as the proper cones resulting from projecting $\K$ on its $x$- and $w$-component, respectively.
Formally, $\K_x = \{\, x \mid \exists w\colon \begin{pmatrix} x \\ w \end{pmatrix} \in \K \,\}$ and $\K_w = \{\, w \mid \exists x\colon \begin{pmatrix} x \\ w \end{pmatrix} \in \K \,\}$.
For those cones, this paper deals with the linear cone-preserving discrete-time switched system
\begin{equation}
\begin{aligned}
	\begin{split} \label{eq:switched-system}
		x(t+1) &= \begin{pmatrix} \mathbf{A}_{\sigma(t)} & \mathbf{B}_{\sigma(t)} \end{pmatrix} \begin{pmatrix} x(t) \\ w(t) \end{pmatrix} \\
		z(t) &= \begin{pmatrix} \mathbf{C}_{\sigma(t)} & \mathbf{D}_{\sigma(t)} \end{pmatrix} \begin{pmatrix} x(t) \\ w(t) \end{pmatrix}
	\end{split}
\end{aligned}
\end{equation}
with state $x(t) \in \K_x$, performance input $w(t) \in \K_w$, performance output $z(t) \in \K_z$, and switching sequence $\sigma$, where $\sigma(t) \in \mathcal{M} = \{\, 1,\dots,m \,\}$ denotes the mode that the switched system is operating in at time $t \geq 0$.
Further, the linear operators defining the system dynamics \eqref{eq:switched-system} are
\begin{equation} \label{eq:switched-system-operators}
	\begin{split}
		\begin{aligned}
			\mathbf{A}_{\sigma(t)}\colon& \K_x \hspace{-1em} {}&&\to \K_x & \qquad
			\mathbf{B}_{\sigma(t)}\colon& \K_w \hspace{-1em} {}&&\to \K_x &\\
			\mathbf{C}_{\sigma(t)}\colon& \K_x \hspace{-1em} {}&&\to \K_z & \qquad
			\mathbf{D}_{\sigma(t)}\colon& \K_w \hspace{-1em} {}&&\to \K_z &
		\end{aligned}
	\end{split}
\end{equation}
and $\begin{pmatrix} \mathbf{A}_{\sigma(t)} \!& \! \mathbf{B}_{\sigma(t)} \end{pmatrix}\!\colon \K \to \K_x$, $\begin{pmatrix} \mathbf{C}_{\sigma(t)} \!&\! \mathbf{D}_{\sigma(t)} \end{pmatrix}\!\colon \K \to \K_z$, consequently.
The system \eqref{eq:switched-system} is \emph{cone-preserving}, \ie
\begin{equation} \label{eq:cone-preserving}
		\begin{pmatrix} x(t) \\ w(t) \end{pmatrix} \in \K \, \Rightarrow \, \big[\, x(t+1) \in \K_x \text{ and } z(t) \in \K_z \,\big],
\end{equation}
which is fulfilled if the operators are satisfying \eqref{eq:switched-system-operators}.
This is not a restriction, but instead a very general property that covers a wide range of systems. In Section~\ref{sec:examples}, we show how two specific and widely used system classes (positive systems and general switched systems), fall into this system setup.

\subsection{Constrained switching} \label{sec:setup-graph}
In many practical scenarios, the switching sequence $\sigma$ in \eqref{eq:switched-system} is not arbitrary, but follows some switching rule.
An established method is to represent the switching rule on $\sigma$ by a finite (possible nondeterministic) automaton, that is a directed graph $\mathcal{G} = (\mathcal{V},\mathcal{E})$ capturing all allowable switching sequences according to the switching rule.
Hereby, $\mathcal{V} = \{\, v_1, \dots, v_\nv \,\}$ is the finite set of nodes and $\mathcal{E} \subseteq \mathcal{V} \times \mathcal{M} \times \mathcal{V}$ is the set of edges with elements $e = (v_i,l,v_j) \in \mathcal{E}$ if there exists an edge in $\mathcal{G}$ from node $v_i$ to node $v_j$ with label $l \in \mathcal{M}$.
The system's mode is $\sigma(t) = l$ when the automaton transitions from node $v_i$ to node $v_j$ along the edge $(v_i,l,v_j)$.
The switched system whose switching sequence is restricted by an automaton as described above can thus be interpreted as a switched system that evolves over time \emph{on} the graph $\mathcal{G}$ by transitioning from a starting node $v(t) = v_i$ at time $t$ along an edge with the associated dynamics $\mathbf{A}_l$, $\mathbf{B}_l$, $\mathbf{C}_l$, $\mathbf{D}_l$, $\sigma(t) = l \in \mathcal{M}$, to an end node $v(t+1) = v_j$ at time $t+1$.
The automaton's current node at time $t$ is indicated by the function
\begin{align} \label{eq:indicator-function}
	\eta_i(t) = \begin{cases}
		1 \quad \text{$v(t) = v_i$, \ie automaton's state at $t$ is $v_i$} \\
		0 \quad \text{otherwise},
	\end{cases} \raisetag{1.5em} %
\end{align}
\ie $\sum_{i=1}^{\nv} v_i \eta_i(t)$ is the automaton's state at time $t$.
Without loss of generality, we assume all nodes $v_i$ to have at least one outgoing edge to ensure that switching sequences of infinite length can be generated.
Note that the automaton formulation also includes arbitrary switching as a special case.

\subsection{System class} \label{sec:setup-system-class}
We can now concisely summarize the in this paper considered system class of \emph{cone-preserving switched system with constrained switching} (CPSS).

\begin{definition}[CPSS] \label{def:CPSS}
	A (linear) cone-preserving switched system with constrained switching is a switched system with cone-preserving dynamics \eqref{eq:switched-system} with \eqref{eq:cone-preserving} described by linear operators~\eqref{eq:switched-system-operators}, corresponding cones $\K$, $\K_z$, and projections $\K_x$, $\K_w$, and a switching rule represented by a finite automaton with graph $\mathcal{G}$ as described above.
\end{definition}
CPSS are the main subject of the work at hand.
While seemingly abstract in their formulation, they allow for a generalization of known system classes, and will lead to an intuitive template for choosing Lyapunov functions, as we will see later.
Next, we define a suitable performance measure for this type of systems, before stating our main result.

\subsection{Performance definition} \label{sec:setup-performance}
In this subsection, we introduce the performance measure used in this paper.
It is based on the fact that given a proper cone $\K'$ and associated inner product, any given $H \in \interior{\K'^*}$ defines a \emph{cone-linear norm} $\norm{y}^H_{\K'} = \inf\limits_{u \in \K'} \{ \innerp{H,u} \, \mid \, \pm y + u \in \K' \}$ with $\norm{y}^H_{\K'} = \innerp{H,y}$ for $y \in \K'$.
See \cite{Seidman2005} for details.

To obtain a for the system suitable performance measure, we use two cone-linear norms: one for the performance input $w$ and one for the performance output $z$.
Picking $H^w \in \interior{\K_w^*}$ and $H^z \in \interior{\K_z^*}$ results in the induced cone-linear performance measure
\begin{equation} \label{eq:performance-def}
	\gamma = \sup\limits_{w \in \mathcal{L}} \frac{\sum_{t=0}^{\infty} \innerp{H^z,z(t)}}{\sum_{t=0}^{\infty} \innerp{H^w,w(t)}},
\end{equation}
with $\mathcal{L} = \bigl\{ w \! \neq \! 0 \mid w(t) \in \K_w, \sum_{t=0}^{\infty} \innerp{H^w,w} < \infty \bigr\}$.
Equivalently, we aim to find the smallest $\gamma > 0$ such that for all $w \in \mathcal{L}$ it holds that $\sum_{t=0}^{\infty} \innerp{H^z,z(t)} < \gamma \sum_{t=0}^{\infty} \innerp{H^w,w(t)}$, formally defined subsequently.

\begin{definition}[$\mathcal{L}$-performance] \label{def:performance}
	The CPSS (Definition~\ref{def:CPSS}) admits $\mathcal{L}$-performance with gain $\gamma > 0$ if it is asymptotically stable and for $x(0) = 0$
	\begin{equation} \label{eq:performance-def2}
		\sum_{t=0}^{\infty} \innerp{H^z,z(t)} < \gamma \sum_{t=0}^{\infty} \innerp{H^w,w(t)} \quad \forall w \in \mathcal{L}.
	\end{equation}
\end{definition}
\vspace*{\belowdisplayskip}%
We remark that this is a very general performance definition that can capture many common performance measures.
Often, similar or equal $H^z$ and $H^w$ are used.
Definition~\ref{def:performance}, however, offers more flexibility.
We substantiate the generality of our setup by showing that two widely used system classes and performance measures are generalized by our setup later in Section~\ref{sec:examples}, namely how $\ell_1$-performance for positive systems and $\ell_2$-performance for general linear switched systems appear naturally.

\section{Main Result} \label{sec:performance}
After defining the system class defined on cones, we present our main results, that is a cone-based performance analysis condition for CPSS.
We will first state our main theorem and follow it up by a discussion and examples.

\begin{theorem}[$\mathcal{L}$-performance of CPSS] \label{thm:performance}
	The CPSS admits $\mathcal{L}$-performance with gain $\gamma > 0$ if there exist $P_i \in \interior{\K_x^*}$, $i = 1 \dots \nv$, such that for all edges $(v_i,l,v_j) \in \mathcal{E}$ of $\mathcal{G}$%
	\begin{equation} \label{eq:performance-cond}
		- \begin{pmatrix}
			 \mathbf{A}_l \!&\!
			 \mathbf{B}_l
		\end{pmatrix}^* \! P_j - \begin{pmatrix}
			 \mathbf{C}_l \!&\!
			 \mathbf{D}_l
		\end{pmatrix}^* \! H^z \!+\! \begin{pmatrix}
			P_i \\
			\gamma H^w
		\end{pmatrix}
		\in \interior{\K^*}.
	\end{equation}
\end{theorem}
\vspace*{\belowdisplayskip}%

\begin{proof}
	The proof uses Lyapunov-based arguments with the Lyapunov function $V$ defined as
	\begin{equation} \label{eq:Lyap}
		V(t) \coloneq V(x(t),t) = \sum_{i=1}^{\nv} \innerp*{P_i,x(t)} \, \eta_i(t),
	\end{equation}
	using the indicator function~\eqref{eq:indicator-function}.
	Observe that by definition of the dual cone \eqref{eq:dual-cone}, $V(x(t),t) > 0$ holds for all $x(t) \in \K_x\backslash\{0\}$ and $P_i \in \interior{\K_x^*}$.
	Note that by \eqref{eq:dual-cone} and linearity of real inner products, \eqref{eq:performance-cond} is equivalent to
	\begin{equation*}\begin{aligned}
		&\innerp*{
			\begin{pmatrix}
				 \mathbf{A}_l & \mathbf{B}_l
			\end{pmatrix}^* P_j,
			\begin{pmatrix}
				x \\ w
			\end{pmatrix}
		} + \innerp*{
			\begin{pmatrix}
				 \mathbf{C}_l & \mathbf{D}_l
			\end{pmatrix}^* H^z,
			\begin{pmatrix}
				x \\ w
			\end{pmatrix}
		} \\
		&- \innerp*{
			\begin{pmatrix}
				P_i \\
				\gamma H^w
			\end{pmatrix},
			\begin{pmatrix}
				x \\ w
			\end{pmatrix}
		} < 0 \qquad \forall \begin{pmatrix} x \\ w \end{pmatrix} \in \K.
	\end{aligned}\end{equation*}
	Replacing the adjoint operators by their counterparts \eqref{eq:adjoint-2elements}, the above equals to
	\begin{equation*}\begin{aligned}
        &\innerp*{
			P_j,
			\begin{pmatrix}
				 \mathbf{A}_l &  \mathbf{B}_l
			\end{pmatrix}
			\begin{pmatrix}
				x \\ w
			\end{pmatrix}
		} + \innerp*{
			H^z,
			\begin{pmatrix}
				 \mathbf{C}_l &  \mathbf{D}_l
			\end{pmatrix}
			\begin{pmatrix}
				x \\ w
			\end{pmatrix}
		} \\
		&- \innerp*{
			\begin{pmatrix}
				P_i \\
				\gamma H^w
			\end{pmatrix},
			\begin{pmatrix}
				x \\ w
			\end{pmatrix}
		} < 0 \qquad \forall \begin{pmatrix} x \\ w \end{pmatrix} \in \K.
	\end{aligned}\end{equation*}
	Consider now an arbitrary time $t$ and the corresponding transition of the automaton's state via the edge $(v_i,l,v_j) \in \mathcal{E}$ of $\mathcal{G}$, \ie a transition in the graph from $v_i$ to $v_j$ with dynamics described by the label $l \in \mathcal{M}$, thus $\sigma(t) = l$.
	Therefore, $\eta_i(t) = 1 = \eta_j(t+1)$ (and $0$ otherwise) and we can now use the system dynamics \eqref{eq:switched-system} and the Lyapunov function \eqref{eq:Lyap} to reformulate above condition into
	\begin{equation*}\begin{aligned}
		&\innerp*{P_j, x(t+1)} + \innerp*{H^z, z(t)}
		- \innerp*{P_i, x(t)} - \gamma \innerp*{H^w, w(t)} \\
		={}& \sum_{i=1}^{\nv} \innerp*{P_i, x(t+1)} \, \eta_j(t+1) - \sum_{i=1}^{\nv} \innerp*{P_i, x(t)} \, \eta_i(t) \\
		&\quad+ \innerp*{H^z, z(t)} - \gamma \innerp*{H^w, w(t)} \\
		={}& V(t+1) - V(t) + \innerp*{H^z, z(t)} - \gamma \innerp*{H^w, w(t)} < 0,
	\end{aligned}\end{equation*}
	or equivalently
		\begin{equation} \label{eq:Lyap-performance-cond}
			V(t+1) - V(t) < \gamma \innerp{H^w,w(t)} - \innerp{H^z,z(t)}.
	\end{equation}
	Next, we show that \eqref{eq:Lyap-performance-cond} implies \eqref{eq:performance-def2}, using arguments along the lines of \cite{Fang2004}.
	Asymptotic stability is obtained directly from \eqref{eq:Lyap-performance-cond} for $w(t)$, $z(t) = 0$ and the fact that $V(x(t),t) > 0$ for all $x(t) \in \K_x\backslash\{0\}$ using standard Lyapunov arguments.
	Recursively applying \eqref{eq:Lyap-performance-cond} results in
    $
		V(t+1) - V(0) < \gamma \sum_{\tau=0}^{t}\innerp{H^w,w(\tau)} - \sum_{\tau=0}^{t}\innerp{H^z,z(\tau)}
    $
	for any $t > 0$. Since $x(0) = 0$ according to Definition~\ref{def:performance}, $V(0) = 0$.
	Letting $t \to \infty$ and observing $0 \leq V(t)$ by definition yields $0 \leq V(\infty) < \gamma \sum_{t=0}^{\infty} \innerp{H^w,w(t)} - \sum_{t=0}^{\infty} \innerp{H^z,z(t)}$.
	Inequality \eqref{eq:performance-def2} and thus $\mathcal{L}$-performance with gain $\gamma$ immediately follows.
\end{proof}

A few notes are in order.
First, \eqref{eq:performance-cond} is a condition that needs to hold for all edges of the graph $\mathcal{G} = (\mathcal{V},\mathcal{E})$.
This is due to the CPSS evolving on the graph, \cf Subsection~\ref{sec:setup-graph}, ensuring that the performance guarantee holds for all admissible switching sequences according to the underlying switching rule.
Further, note that the generality of Theorem~\ref{thm:performance} allows to recover known results from the literature.
For the two special cases of positive systems and general switched linear systems with constrained switching, we show in the next section how to recover known results from~\eqref{eq:performance-cond}.
Additionally, Theorem~\ref{thm:performance} implicitly contains stability: In the absence of a performance measure, \ie $H^w$, $H^z = 0$, \eqref{eq:performance-cond} becomes a condition for asymptotic stability.

\begin{remark}[Lyapunov function]
    Equation \eqref{eq:Lyap} with the $P_i$ of \eqref{eq:performance-cond} constitutes a so-called \emph{switched Lyapunov function}.
	Those are widely used for switched systems \cite{Lin2006}, consist of one scenario-specific Lyapunov function per node $v_i \in \mathcal{V}$ of $\mathcal{G}$, and switch between them whenever the automaton transitions to the next node.
	Especially for constrained switching they are known to increase the feasibility compared to a non-switched Lyapunov function, which is obtained for $P_i = P$.
	Equation~\eqref{eq:Lyap} further poses a structured Lyapunov function candidate based on the cone-preserving structure of the CPSS.
    This yields insights and a systematic approach to Lyapunov function selection for those systems.
	We recover commonly used types of Lyapunov functions in the next section.
\end{remark}

\section{Examples} \label{sec:examples}
In this section we illustrate our main result by applying Theorem~\ref{thm:performance} to two special cases of CPSSs: a positive switched system and a general switched system.
Both system classes are widely used and investigated in the literature.
For both, we derive a performance result based on Theorem~\ref{thm:performance}, partially recovering known results from the literature.
In doing so, we show the generality of our main result.
Additionally, the derived result for positive switched systems with constrained switching is to the best of our knowledge unknown to the literature so far.

\subsection{Positive systems} \label{sec:examples-positive-system}
In this subsection, we show that the class of positive switched systems with constrained switching (PSS) is a special case of Definition~\ref{def:CPSS}.
Further, we evaluate \eqref{eq:performance-cond} for PSS and thereby arrive at a novel $\ell_1$-performance condition for PSS.
So far, only $\ell_1$-performance results for non-switched positive systems are known \cite{Chen2013}.
Positive systems are characterized by the fact that state, input, and output variables are always nonnegative \cite{Rantzer2018}.
Let consequently $\begin{pmatrix} x(t) \\ w(t) \end{pmatrix} \in \K = \K_x \times \K_w = \mathbb{R}^n_{\geq 0} \times \mathbb{R}^q_{\geq 0} = \K^*$ and $z(t) \in \K = \mathbb{R}^r_{\geq 0} = \K^*_z$, \ie the cones $\K$ and $\K_z$ are self-dual.
The equipped inner product for all those cones is $\innerp{x,y} = x^\top y$.
Henceforth, the operators $ \mathbf{A}_l$, $ \mathbf{B}_l$, $ \mathbf{C}_l$, $ \mathbf{D}_l$ are matrix-vector multiplications, where the matrices have nonnegative entries only \cite{Rantzer2018}.

\begin{definition}[Positive switched system]
	A (linear) positive switched system with constrained switching (PSS) is given by the dynamics
    \begin{equation}\begin{aligned} \label{eq:positive-system}
		\begin{split}
			x(t+1) &= A_{\sigma(t)} x(t) + B_{\sigma(t)} w(t) \\
			z(t) &= C_{\sigma(t)} x(t) + D_{\sigma(t)} w(t)
		\end{split}
	\end{aligned}\end{equation}
	with $x(t) \in \mathbb{R}^n_{\geq 0}$, $w(t) \in \mathbb{R}^q_{\geq 0}$, $z(t) \in \mathbb{R}^r_{\geq 0}$,
	$A_{\sigma(t)} \in \mathbb{R}^{n \times n}_{\geq 0}$, $B_{\sigma(t)} \in \mathbb{R}^{n \times q}_{\geq 0}$, $C_{\sigma(t)} \in \mathbb{R}^{r \times n}_{\geq 0}$, $D_{\sigma(t)} \in \mathbb{R}^{r \times q}_{\geq 0}$, and switching sequence $\sigma$ with a switching rule represented by a finite automaton with graph $\mathcal{G}$, as described in Subsection~\ref{sec:setup-graph}.
\end{definition}

To write the PSS \eqref{eq:positive-system} in the form of a CPSS \eqref{eq:switched-system}, define the operators and their adjoints, given by the transpose, as
\begin{equation}
	\begin{split}
		\begin{aligned} \label{eq:positive-system-operators}
			\begin{pmatrix}  \mathbf{A}_{\sigma(t)} &  \mathbf{B}_{\sigma(t)} \end{pmatrix}\colon \begin{pmatrix} x \\ w \end{pmatrix} &\mapsto A_{\sigma(t)} x + B_{\sigma(t)} w && \\
			\begin{pmatrix}  \mathbf{C}_{\sigma(t)} &  \mathbf{D}_{\sigma(t)} \end{pmatrix}\colon \begin{pmatrix} x \\ w \end{pmatrix} &\mapsto C_{\sigma(t)} x + D_{\sigma(t)} w
		\end{aligned}
	\end{split}
\end{equation}
\begin{equation}
    \begin{split}
		\begin{aligned} \label{eq:positive-system-operators-adjoint}
			\begin{pmatrix}  \mathbf{A}_{\sigma(t)} &  \mathbf{B}_{\sigma(t)} \end{pmatrix}^*\colon p &\mapsto \begin{pmatrix} A_{\sigma(t)}^\top p \\ B_{\sigma(t)}^\top p \end{pmatrix} && \\
			\begin{pmatrix}  \mathbf{C}_{\sigma(t)} &  \mathbf{D}_{\sigma(t)} \end{pmatrix}^*\colon h &\mapsto \begin{pmatrix} C_{\sigma(t)}^\top h \\ D_{\sigma(t)}^\top h \end{pmatrix}.
		\end{aligned}
	\end{split}
\end{equation}
A commonly used performance measure for positive systems is given by the $\ell_1$-gain \cite{Chen2013}.
For $H^w = h^w = \mathsf{1} \in \interior{\K_w^*}$ and $H^z = h^z = \mathsf{1} \in \interior{\K_z^*}$, we have
\begin{equation} \label{eq:performance-l1}
	\gamma = \sup\limits_{w \in \ell_1} \frac{\sum_{t=0}^{\infty} \mathsf{1}^\top z(t)}{\sum_{t=0}^{\infty} \mathsf{1}^\top{w(t)}} = \sup\limits_{w \in \ell_1} \frac{\norm{z}_1}{\norm{w}_1}
\end{equation}
and $\mathcal{L} = \ell_1 = \{\, w \neq 0 \mid w(t) \in \mathbb{R}^q_{\geq 0}, \sum_{t=0}^{\infty} \mathsf{1}^\top w(t) < \infty \,\}$, recovering in the $\ell_1$-induced norm as one natural performance measure for positive systems.
Since $\interior{\K_x^*} = \mathbb{R}^n_{> 0}$, $V(x(t),t) = \sum_{i=1}^{\nv} p_i^\top x(t) \eta_i(t) \quad \text{with} \quad p_i \in \mathbb{R}^n_{>0}$
qualifies as a (piecewise) linear Lyapunov function according to \eqref{eq:Lyap}.
(Piecewise) linear Lyapunov functions are known to be a suitable Lyapunov function type for positive (switched) systems \cite{Liu2009,Fornasini2012,Chen2013,Rantzer2018}.
Using \eqref{eq:positive-system-operators-adjoint}, the performance condition \eqref{eq:performance-cond} for $\ell_1$-performance then reads as
\begin{equation*}\begin{aligned}
	&-\begin{pmatrix}
		A_l^\top p_j \\
		B_l^\top p_j
	\end{pmatrix} - \begin{pmatrix}
		C_l^\top h^z \\
		D_l^\top h^z
	\end{pmatrix} + \begin{pmatrix}
		p_i \\
		\gamma h^w
	\end{pmatrix} \in \interior{\K^*} \\
	\Leftrightarrow{}& - \begin{pmatrix}
		A_l^\top p_j - p_i + C_l^\top \mathsf{1} \\
		B_l^\top p_j - \gamma \mathsf{1} + D_l^\top \mathsf{1}
	\end{pmatrix} \in \interior{\K^*} = \mathbb{R}^n_{> 0} \times \mathbb{R}^q_{> 0}.
\end{aligned}\end{equation*}
The subsequent performance result for PSS follows.

\begin{corollary}[$\ell_1$-performance of PSS] \label{cor:performance-positive-systems}
	The PSS admits $\ell_1$-performance with gain $\gamma > 0$ if there exists $p_i \in \mathbb{R}^n_{> 0}$, $i = 1 \dots \nv$, such that for all edges $(v_i,l,v_j) \in \mathcal{E}$
	\begin{subequations} \label{eq:performance-cond-positive-system}
		\begin{align}
			A_l^\top p_j - p_i + C_l^\top \mathsf{1} \, &\in \mathbb{R}^n_{<0} \label{eq:performance-cond-positive-systems-a}\\
			B_l^\top p_j - \gamma \mathsf{1} + D_l^\top \mathsf{1} \, &\in \mathbb{R}^q_{<0}.\label{eq:performance-cond-positive-systems-b}
		\end{align}
	\end{subequations}
\end{corollary}
\vspace*{\belowdisplayskip}%

Corollary~\ref{cor:performance-positive-systems} poses a novel result that allows to analyze $\ell_1$-performance for positive switched systems for constrained switching.
It further generalizes some results known from the literature:
$\ell_1$-performance analysis results for positive non-switched systems \cite[Theorem~2]{Chen2013}, \cite[Theorem~2]{Zhu2024} for discrete-time systems and \cite[Theorem~4]{Shen2017} for continuous-time cone-preserving systems; stability results for positive switched systems, \ie only \eqref{eq:performance-cond-positive-systems-a} for $C_l = 0$, with arbitrary switching using switched \cite[Theorem~1]{Liu2009}, \cite[Theorem~8.3]{Benzaouia2012} and non-switched Lyapunov functions \cite[Theorem~1]{Fornasini2012}, \ie $p_i = p$ for all $i$.
Note again that arbitrary switching is a special case of constrained switching, that can be represented by a specific automaton that allows all switching sequences.

Conditions~\eqref{eq:performance-cond-positive-system} are solved efficiently using linear programming, since they are element-wise linear inequalities.
A numerical example thereof is given below.

\subsection{Numerical example for positive systems} \label{sec:examples-positive-system-numerics}
\begin{figure}
	\centering
	\begin{tikzpicture}
		\node[vertexFill] (v1) at (0,0.5) {$v_1$};
		\node[vertexFill] (v2) at (1.75,1.25) {$v_2$};
		\node[vertexFill] (v3) at (3,0) {$v_3$};
		\node[vertexFill] (v4) at (1.5,0) {$v_4$};

		\draw[edge] (v1) to[bend left] node[above left, pos=0.5] {\footnotesize$m_1$} (v2);
		\draw[edge] (v2) to[bend left] node[above right, pos=0.5] {\footnotesize$m_3$} (v3);
		\draw[edge] (v3) to[loop right] node[above, pos=0.5] {\footnotesize$m_3$} (v3);
		\draw[edge] (v3) to[bend left] node[above, pos=0.5] {\footnotesize$m_2$} (v4);
		\draw[edge] (v4) to[bend left] node[above, pos=0.5] {\footnotesize$m_3$} (v3);
		\draw[edge] (v4) to[bend left] node[below, pos=0.5] {\footnotesize$m_1$} (v1);
	\end{tikzpicture}
	\caption{Switching rule graph for the virus mitigation system.}
	\label{fig:example}
\end{figure}
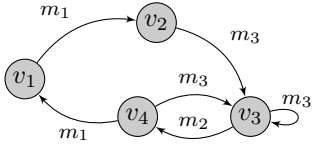

To illustrate the performance result for PSS, we present a simplified virus mitigation system, that we will analyze for its $\ell_1$-performance.
It consist of three countries A, B, and C with common borders, state $x_i$ representing the number of infections in country $i \in \{\, \text{A}, \text{B}, \text{C} \,\}$, and each time step representing one week.
Each of them has a infection rate $\lambda_i$ with $\lambda_\mathrm{A} = 1.2$, $\lambda_\mathrm{B} = 1.4$, $\lambda_\mathrm{C} = 1.4$, for example due to different population density, and can employ a quarantine rate $k_i$.
Further, every country is subject to some spontaneous infections of various origin, modeled by the disturbance input $w$ and the constant factor $100$.
The three different countries pursue different virus mitigation strategies.
    Country A is quarantining a constant ratio of the infected population ($k_\mathrm{A} = 0.5$).
    Country B has a more involved strategy with 3 measure levels $m_1$, $m_2$, $m_3$: For at least one week, the borders to both A and C are closed, and infected persons are consequently quarantined ($k_\mathrm{B}(m_3) = 1$). After that, the measures may be partially lowered, where the borders remain closed but nobody is quarantined ($k_\mathrm{B}(m_2) = 0$). Then, the borders are either reopened for at least two weeks ($m_1$, $k_\mathrm{B}(m_1) = 0$), or full measures are in place again ($m_3$).
    Country C coordinates their measures with B. No borders are closed, but for measure levels $m_2$ and $m_3$, infected people are partially quarantined ($k_\mathrm{C}(m_2, m_3) = 0.6$).
Putting the above in our framework leads to a CPSS with a switching rule represented by the graph given in Figure~\ref{fig:example}.
As performance measure, we consider the number of total infected people, that is $\norm{x}_1$, \ie the performance output is $z(t) = \mathsf{1}^\top x(t)$.
Denoting $\Lambda = \mathrm{diag}(\lambda_\mathrm{A},\lambda_\mathrm{B},\lambda_\mathrm{C})$ and $K = \mathrm{diag}(k_\mathrm{A},k_\mathrm{B},k_\mathrm{C})$, where $\mathrm{diag}(\cdot)$ is the diagonal matrix of the arguments, the virus mitigation system is described by the positive switched system
\begin{equation*}
    \begin{aligned}
		x(t+1) &= \underbrace{(\Lambda - K(m_i) + S(m_i))}_{\eqcolon A_{\sigma(t)}} x(t) + B w(t) \\
        z(t) &= C x(t)
    \end{aligned}
\end{equation*}
with $\sigma(t) \in \{\, m_1, m_2, m_3 \,\}$, and non-switched $B = 100 \cdot \mathsf{1}$ and $C = \mathsf{1}^\top$.
Here, the virus spread through borders between the countries is described by the matrix $S$ with
\[
	S(m_1) \!=\! s \! \begin{pmatrix}
		-2 \!\! & 1 & 1 \\
		1 & \!\! -2 \!\! & 1 \\
		1 & 1 & \!\! -2
	\end{pmatrix}\!, ~ S(m_2, m_3) \!=\! s \! \begin{pmatrix}
		-1 \!\! & 0 \! & 1 \\
		0 & 0 \! & 0 \\
		1 & 0 \! & \!\! -1
	\end{pmatrix}\!,
\]
where $s = 0.4$.
Observe first that for arbitrary switching (or equivalently: no measures, measure level $m_1$) the system is unstable, which can easily be seen by the fact that an eigenvalue of $A_{m_1}$ is larger than $1$.
Analyzing this system using Corollary~\ref{cor:performance-positive-systems}, the obtained $\ell_1$-performance is $\gamma = \SI{9451}{}$, which is the total number of infections in relation to the number of spontaneous infections, \cf \eqref{eq:performance-l1}.

Country C is thinking about deploying a more rigorous quarantine strategy.
If $k_\mathrm{C}(m_2,m_3) = 0.8$, the $\ell_1$-performance is improved to $\gamma = \SI{4586}{}$.
On the other hand, if Country B fails to quarantine all infected ($k_\mathrm{B}(m_1) = 0.85$), we have $\gamma = \SI{109145}{}$, drastically increasing the overall infections.

\subsection{General switched systems} \label{sec:examples-semidef-cone}

In this section we show that Theorem~\ref{thm:performance} also covers known results for switched systems. To this end, consider the general (linear) switched system with constrained switching (GSS)
\begin{equation}\begin{aligned} \label{eq:switched-system-general}
	\begin{split}
		x(t+1) &= A_{\sigma(t)} x(t) + B_{\sigma(t)} w(t)\\
		z(t) &= C_{\sigma(t)} x(t) + D_{\sigma(t)} w(t)
	\end{split}
\end{aligned}\end{equation}
with $x(t) \in \mathbb{R}^n$, $w(t) \in \mathbb{R}^q$, $z(t) \in \mathbb{R}^r$, and matrices $A_{\sigma(t)} \in \mathbb{R}^{n \times n}$, $B_{\sigma(t)} \in \mathbb{R}^{n \times q}$, $C_{\sigma(t)} \in \mathbb{R}^{r \times n}$, and $D_{\sigma(t)} \in \mathbb{R}^{r \times q}$ of suitable dimensions.
No further assumptions on the system \eqref{eq:switched-system-general} are made except that it follows a switching rule represented by an automaton and an underlying graph $\mathcal{G}$, as described in Subsection~\ref{sec:setup-graph}.

To be able to apply our results, it is necessary to lift the system dynamics in \eqref{eq:switched-system-general} onto the so-called \emph{semi-definite cone} \cite{Boyd2004}, equipped with the inner product $\innerp{H,X} = \mathrm{tr}(HX)$.
Such representations are commonly used when describing the evolution of the covariance of the system state in the stochastic setting \cite{Skelton2017}.
For our purposes it is sufficient to introduce the variables
\[
Y(t) \coloneq \begin{pmatrix}
    X(t) \!&\! W_1(t)\\W_1(t)^\top \!& \!W_2(t)
\end{pmatrix}\in\mathbb{R}^{(n+q)\times{}(n+q)},\,Z(t)\in\mathbb{R}^{r\times{}r}.
\]
We now set $\mathcal{K}$ to be the set of $(n+q)\times{}(n+q)$ positive semi-definite matrices, and $\mathcal{K}_z$ to be the set of $r\times{}r$ positive semi-definite matrices. Note that in a slight abuse of notation, the variables $X(t)$, $(W_1(t),W_2(t))$, and $Z(t)$ will play the roles of $x(t)$, $w(t)$, and $z(t)$ respectively, in the definition of the CPSS according to Definition~\ref{def:CPSS} (we will use these matrix variables for this purpose throughout the rest of the subsection to avoid confusion with the variables in \eqref{eq:switched-system-general}). 

Consider now the linear operators $\begin{pmatrix}  \mathbf{A}_{\sigma(t)} &  \mathbf{B}_{\sigma(t)} \end{pmatrix}\colon\mathcal{K}\rightarrow{}\mathcal{K}_x$ and $\begin{pmatrix}  \mathbf{C}_{\sigma(t)} & \mathbf{D}_{\sigma(t)} \end{pmatrix}\colon\mathcal{K}\rightarrow{}\mathcal{K}_z$ defined according to
\begin{equation*}
    \begin{aligned}
        Y &\mapsto
        \begin{pmatrix}A_{\sigma(t)} \!&\! B_{\sigma(t)} \end{pmatrix}
        Y
        \begin{pmatrix}A_{\sigma(t)} & B_{\sigma(t)} \end{pmatrix}^\top \\
        Y &\mapsto
        \begin{pmatrix}C_{\sigma(t)} \!&\! D_{\sigma(t)} \end{pmatrix}
        Y
        \begin{pmatrix}C_{\sigma(t)} & D_{\sigma(t)}\end{pmatrix}^\top.
    \end{aligned}
\end{equation*}
We are now ready to define the lifted counterpart to \eqref{eq:switched-system-general}.
\begin{definition}[Lifted system]
	The cone-preserving lifted system representation of \eqref{eq:switched-system-general} is given by the dynamics
	\begin{equation}\begin{aligned}
		\begin{split} \label{eq:switched-system-lifted}
			X(t+1) &=  \begin{pmatrix}
			    \mathbf{A}_{\sigma(t)} &  \mathbf{B}_{\sigma(t)} 
                \end{pmatrix}
                \begin{pmatrix}
                    X(t)&W_1(t)\\W_1(t)^\top{}&W_2(t)
                \end{pmatrix} \\
			Z(t) &=  \begin{pmatrix}
			    \mathbf{C}_{\sigma(t)} &  \mathbf{D}_{\sigma(t)} 
                \end{pmatrix}
                \begin{pmatrix}
                    X(t)&W_1(t)\\W_1(t)^\top{}&W_2(t)
                \end{pmatrix},
		\end{split}
	\end{aligned}\end{equation}
	with state $X(t) \in \K_x$, performance input $W(t) \in \K_w$ and output $Z(t) \in \K_z$, operators $(\mathbf{A}_{\sigma(t)},\mathbf{B}_{\sigma(t)},\allowbreak\mathbf{C}_{\sigma(t)},\mathbf{D}_{\sigma(t)})$ as above, and the switching sequence $\sigma$, retaining the switching rule and underlying graph $\mathcal{G}$ from system \eqref{eq:switched-system-general}.
\end{definition}

Observe in particular that whenever the matrices
$Y$
are rank one, we may recover the dynamics in \eqref{eq:switched-system-general} according to
$Y=
\begin{pmatrix}
    x(t)\\w(t)
\end{pmatrix}
\begin{pmatrix}
    x(t)\\w(t)
\end{pmatrix}^{\top}\!\!\!\!\!$$Z(t)=z(t)z(t)^\top{}.$
Hence, the trajectories of \eqref{eq:switched-system-general} correspond to the rank one trajectories of the lifted GSS in \eqref{eq:switched-system-lifted}.
Therefore, any performance guarantees that we can obtain for the lifted system will also hold for the original GSS.
We will now see that an application of Theorem~\ref{thm:performance} recovers exactly the known $\ell_2$-performance guarantees for \eqref{eq:switched-system-general} from \cite{Seidel2024b}, and further generalizes the conditions from \cite{Linsenmayer2017} for stability analysis.
In order to apply Theorem~\ref{thm:performance} to the lifted GSS we must define some cone-linear norms via $H^w\in\interior{\mathcal{K}_w^*}$ and $H^z\in\interior{\mathcal{K}_z^*}$ as outlined in Subsection~\ref{sec:setup-performance}.
Note that the semi-definite cone is self-dual, and we may therefore set $H^z$ to be any appropriately dimensioned positive definite matrix.
Since
$
\mathcal{K}_w=\left\{\,(W_1,W_2)\mid\exists{}X\succeq{}0\colon Y \succeq{}0\,\right\}
$
it follows from the definition of the dual cone that
$\interior{\mathcal{K}_w^*}=\left\{\,(0,W)\mid
    W\succ{}0\,\right\}$.
If we then set $H^w=(0,I)\in\interior{\mathcal{K}_w^*}$ and $H^z=I\in\interior{\mathcal{K}^*_z}$, the induced cone-linear performance measure in \eqref{eq:performance-def} becomes
$
	\gamma = \sup\limits_{W_2 \in \mathcal{L}} \frac{\sum_{t=0}^{\infty} \mathrm{tr}(Z(t))}{\sum_{t=0}^{\infty} \mathrm{tr}(W_2(t))},
$
where $\mathcal{L} = \bigr\{\, W_2 \neq 0 \mid W_2(t)\succeq{}0, \sum_{t=0}^{\infty} \mathrm{tr}(W_2(t)) < \infty \,\bigl\}$.
Observe in particular that along the rank one trajectories of \eqref{eq:switched-system-lifted} (which correspond to the trajectories of the original GSS), we have the correspondences $W_2(t)=w(t)w(t)^\top{}$ and $Z(t)=z(t)z(t)^\top$. It follows that along those trajectories
\vspace{-0.1em}
\[
\frac{\sum_{t=0}^{\infty} \mathrm{tr}(Z(t))}{\sum_{t=0}^{\infty} \mathrm{tr}(W_2(t))}=\frac{\sum_{t=0}^{\infty}z(t)^\top{}z(t)}{\sum_{t=0}^\infty{}w(t)^\top{}w(t)}=\left(\frac{\Vert{z}\Vert_{\ell_2}}{\Vert{}w\Vert_{\ell_2}}\right)^2,
\vspace{-0.0em}
\]
using the cyclic property of the trace.
Therefore, $\gamma$ not only bounds the induced cone-linear performance of the lifted system, but also upper bounds \textit{the square} of the induced $\ell_2$-norm of the original GSS.
This is why $\gamma$ rather than $\gamma^2$ appears in the LMIs to follow, in contrast to other works in the literature \cite{Fang2004,Seidel2024b}.
We now apply Theorem~\ref{thm:performance}. Noting that
\vspace{-0.3em}
\begin{equation}
\begin{aligned}
	\begin{split} \label{eq:semidef-cone-operators-adjoint}
		\begin{pmatrix}  \mathbf{A}_{\sigma(t)} &  \mathbf{B}_{\sigma(t)} \end{pmatrix}^* P
		&= \begin{pmatrix}
			A_{\sigma(t)}^\top \\ B_{\sigma(t)}^\top
		\end{pmatrix} P \begin{pmatrix}
			A_{\sigma(t)} & B_{\sigma(t)}
		\end{pmatrix} \\
		\begin{pmatrix}  \mathbf{C}_{\sigma(t)} &  \mathbf{D}_{\sigma(t)} \end{pmatrix}^* H
		&= \begin{pmatrix}
			C_{\sigma(t)}^\top \\ D_{\sigma(t)}^\top
		\end{pmatrix} H \begin{pmatrix}
			C_{\sigma(t)} & D_{\sigma(t)}
		\end{pmatrix},
	\end{split}
\end{aligned}
\vspace{-0.2em}
\end{equation}
it follows that \eqref{eq:performance-cond} becomes
\vspace{-0.1em}
\begin{equation*}
\begin{aligned}
	-\begin{pmatrix}
		A_l^\top \\ B_l^\top
	\end{pmatrix} P_j \begin{pmatrix}
		A_l & B_l
	\end{pmatrix} &- \begin{pmatrix}
		C_l^\top \\ D_l^\top
	\end{pmatrix} \begin{pmatrix}
		C_l & D_l
	\end{pmatrix}\\
    &+ \mathrm{diag}(P_i, \gamma I) \in \interior{\K^*}.
\end{aligned}
\vspace{-0.1em}
\end{equation*}
Since $\K$ is the semi-definite cone, which is self-dual, the above can be rewritten as
\begin{equation}\begin{aligned} \label{eq:performance-cond-semidef-cone}
	\begin{pmatrix}
		A_l^\top P_j A_l - P_i + C_l^\top C_l& A_l^\top P_j B_l + C_l^\top D_l \\
		B_l^\top P_j A_l + D_l^\top C_l & B_l^\top P_j B_l + D_l^\top D_l - \gamma I
	\end{pmatrix} \prec 0. \raisetag{0.5em}
\end{aligned}\end{equation}
The following is then an immediate consequence of the observation that the trajectories of \eqref{eq:switched-system-general} correspond to the rank one trajectories of \eqref{eq:switched-system-lifted}.

\begin{corollary}[$\ell_2$-performance of GSS] \label{cor:performance-general-switched-systems}
	The GSS admits $\ell_2$-performance with gain $\gamma > 0$ if there exists $P_i \succ 0$, $i = 1 \dots \nv$, such that \eqref{eq:performance-cond-semidef-cone} holds for all edges $(v_i,l,v_j) \in \mathcal{E}$.
\end{corollary}

Corollary~\ref{cor:performance-general-switched-systems} recovers the known result for $\ell_2$-performance of GSS and the fact that suitable Lyapunov function candidates for GSS are positive definite matrices, as given by \cite[Theorem~1]{Seidel2024b}.
Therein, some reformulations to the linear matrix inequalities \eqref{eq:performance-cond-semidef-cone} are made to facilitate controller synthesis.
From the last equation of the respective proof in~\cite{Seidel2024b}, \eqref{eq:performance-cond-semidef-cone} can be recovered.
Corollary~\ref{cor:performance-general-switched-systems} further generalizes \cite[Corollary~9]{Linsenmayer2017} for stability analysis of GSS. For a numerical example for $\ell_2$-performance of a general linear switched system with constrained switching, we refer to \cite{Seidel2024b}.

\section{Conclusion} \label{sec:conclusion}
For cone-preserving switched systems with constrained switching, we utilized cone properties and arguments to pose graph-based conditions for performance analysis.
The presented unifying perspective provides a template for appropriate Lyapunov function selection and our performance conditions recovers several existing results from the literature.
Specifically, we derived an $\ell_1$-performance analysis result for positive switched systems with constrained switching, a result that was a gap in literature so far.
We showed how our results applly to general switched systems, illustrating the generality of the developed theory.
Future research includes controller synthesis and verifying the cone-preserving property.

\bibliographystyle{ieeetr}
\bibliography{literature.bib}

\begin{thebibliography}{10}

\bibitem{Lin2009}
H.~Lin and P.~J. Antsaklis, ``Stability and {S}tabilizability of {S}witched
  {L}inear {S}ystems: {A} {S}urvey of {R}ecent {R}esults,'' {\em {IEEE} Trans.
  Autom. Control}, vol.~54, no.~2, pp.~308--322, 2009.

\bibitem{Rantzer2018}
A.~Rantzer and M.~E. Valcher, ``A {T}utorial on {P}ositive {S}ystems and
  {L}arge {S}cale {C}ontrol,'' in {\em Proc. 57th Conf. Decision and Control
  (CDC)}, pp.~3686--3697, 2018.

\bibitem{Chen2013}
X.~Chen, J.~Lam, P.~Li, and Z.~Shu, ``l1-induced norm and controller synthesis
  of positive systems,'' {\em Automatica}, vol.~49, no.~5, pp.~1377--1385,
  2013.

\bibitem{Fornasini2012}
E.~Fornasini and M.~E. Valcher, ``Stability and {S}tabilizability {C}riteria
  for {D}iscrete-{T}ime {P}ositive {S}witched {S}ystems,'' {\em {IEEE} Trans.
  Autom. Control}, vol.~57, no.~5, pp.~1208--1221, 2012.

\bibitem{Liu2009}
X.~Liu, ``Stability {A}nalysis of {S}witched {P}ositive {S}ystems: A {S}witched
  {L}inear {C}opositive {L}yapunov {F}unction {M}ethod,'' {\em {IEEE} Trans.
  Circuits and Systems II: Express Brief}, vol.~56, no.~5, pp.~414--418, 2009.

\bibitem{Shen2017}
J.~Shen and J.~Lam, ``Input–output gain analysis for linear systems on
  cones,'' {\em Automatica}, vol.~77, pp.~44--50, Mar. 2017.

\bibitem{Forni2017}
F.~Forni, R.~Jungers, and R.~Sepulchre, ``Path-complete positivity of switching
  systems,'' {\em IFAC-PapersOnLine}, vol.~50, no.~1, pp.~4558--4563, 2017.

\bibitem{Vreman2022b}
N.~Vreman, P.~Pazzaglia, V.~Magron, J.~Wang, and M.~Maggio, ``Stability of
  {L}inear {S}ystems {U}nder {E}xtended {W}eakly-{H}ard {C}onstraints,'' {\em
  {IEEE} Control Systems Letters}, vol.~6, pp.~2900--2905, 2022.

\bibitem{Seidel2024b}
M.~Seidel, S.~Lang, and F.~Allgöwer, ``On $\ell2$-performance of weakly-hard
  real-time control systems,'' {\em European Journal of Control}, vol.~80,
  p.~101056, 2024.

\bibitem{An2024}
S.~An, F.~Wu, J.~Lian, and D.~Wang, ``Graph-{B}ased {R}estricted and
  {A}rbitrary {S}witching for {S}witched {P}ositive {S}ystems via a {W}eak
  {CLCLF},'' {\em {IEEE} Trans. Cybernetics}, vol.~54, no.~8, pp.~4454--4463,
  2024.

\bibitem{Benzaouia2012}
A.~Benzaouia, {\em Stability and {S}tabilization of {P}ositive {S}witching
  {L}inear {D}iscrete-{T}ime {S}ystems}, pp.~195--216.
\newblock Springer London, 2012.

\bibitem{Shen2010}
J.~Shen and J.~Hu, ``Stability of switched linear systems on cones: A
  generating function approach,'' in {\em Proc. 49th Conf. Decision and Control
  (CDC)}, pp.~420--425, 2010.

\bibitem{Bundfuss2009}
S.~Bundfuss and M.~Dür, ``Copositive {L}yapunov functions for switched systems
  over cones,'' {\em Systems \& Control Letters}, vol.~58, no.~5, pp.~342--345,
  2009.

\bibitem{Zhu2024}
B.~Zhu, J.~Lam, J.~Shen, Y.~Cui, X.~Lu, and K.-W. Kwok, ``Input–{O}utput
  {G}ain {A}nalysis of {L}inear {D}iscrete-{T}ime {S}ystems {W}ith {C}one
  {I}nvariance,'' {\em {IEEE} Trans. Autom. Control}, vol.~69, no.~12,
  pp.~8751--8757, 2024.

\bibitem{Boyd2004}
S.~Boyd and L.~Vandenberghe, {\em Convex {O}ptimization}.
\newblock Cambridge University Press, 2004.

\bibitem{Seidman2005}
T.~I. Seidman, H.~Schneider, and M.~Arav, ``Comparison theorems using general
  cones for norms of iteration matrices,'' {\em Linear Algebra and its
  Applications}, vol.~399, pp.~169--186, 2005.

\bibitem{Fang2004}
L.~Fang, H.~Lin, and P.~Antsaklis, ``Stabilization and performance analysis for
  a class of switched systems,'' in {\em Proc. 43rd Conf. Decision and Control
  (CDC)}, vol.~3, pp.~3265--3270, 2004.

\bibitem{Lin2006}
H.~Lin and P.~J. Antsaklis, ``Switching {S}tabilization and l2 {G}ain
  {P}erformance {C}ontroller {S}ynthesis for {D}iscrete-{T}ime {S}witched
  {L}inear {S}ystems,'' in {\em Proc. 45th Conf. Decision and Control (CDC)},
  2006.

\bibitem{Skelton2017}
R.~E. Skelton, T.~Iwasaki, and K.~M. Grigoriadis, {\em Unified Algebraic
  Approach to Control Design}.
\newblock CRC Press LLC, 2017.

\bibitem{Linsenmayer2017}
S.~Linsenmayer and F.~Allgöwer, ``Stabilization of networked control systems
  with weakly hard real-time dropout description,'' in {\em Proc. 56th Conf.
  Decision and Control (CDC)}, pp.~4765--4770, 2017.

\end{thebibliography}
\end{document}